\documentclass[prb,bibnotes,twocolumn]{revtex4}
\usepackage{amsmath}

\usepackage{graphicx}
\usepackage{hyperref}
\usepackage{graphicx}
\usepackage{amsmath,amssymb}
\usepackage{float}
\usepackage{bm}

\begin{document}
\draft
\def\ds{\displaystyle}
\title{Coexistence of extended and localized states in one-dimensional non-Hermitian Anderson model} 
\author{Cem Yuce$^{1}$, Hamidreza Ramezani$^2$}
%\email{cyuce@eskisehir.edu.tr}
\address{$^1$ Department of Physics, Eskisehir Technical University, Eskisehir, Turkey\\
$^2$ Department of Physics and Astronomy, University of Texas Rio Grande Valley, Edinburg, TX 78539, USA }

\date{\today}
%\pacs{ }
\begin{abstract}

In one-dimensional Hermitian tight-binding models, mobility edges separating extended and localized states can appear in the presence of properly engineered quasi-periodical potentials and coupling constants. On the other hand, mobility edges don't exist in a one-dimensional Anderson lattice since localization occurs whenever a diagonal disorder through random numbers is introduced. Here, we consider a nonreciprocal non-Hermitian lattice and show that the coexistence of extended and localized states appears with or without diagonal disorder in the topologically nontrivial region. We discuss that the mobility edges appear basically due to the boundary condition sensitivity of the nonreciprocal non-Hermitian lattice.
\end{abstract}
\maketitle

\section{Introduction}

Anderson localization (AL), a well-understood fundamental problem in condensed matter, is the absence of diffusion of waves in a disordered medium due to interference of waves \cite{AL0}. Specifically in AL, all states are exponentially localized in the presence of any disorder in one and two-dimensional Anderson model at which a random disordered on-site potential is introduced. On the other hand for weak disorder if the localization length is bigger than the system size then the system behaves as it is delocalized. In three dimensions,  we would have a mobility edge separating localized and extended states. On contrary to the one dimensional (1D) Anderson model, in the Aubry-Andr\'e model in which its disorder is modeled as a quasi-periodic on-site potential depending on the strength of incommensurate potential, all states are localized or delocalized \cite{AL1}. This means that the system can undergo a metal-insulator transition even in 1D. However, this transition is sharp, i.e. all single-particle eigenstates in the spectrum suddenly become exponentially localized above a threshold level of disorder. In both cases, localized and extended states generally do not coexist since non of these models possess a mobility edge in 1D, i.e.,
critical energy separates localized and delocalized energy eigenstates. Recent studies show that the transition is not sharp beyond the one-dimensional Aubry-Andr\'e model with correlated disorder and hopping amplitudes. It was shown that an intermediate regime characterized by the coexistence of localized and extended states at different energies may occur \cite{AL3,AL4,AL4uneki,AL4uneki2,AL4uneki3,AL4uneki4}. The theoretical findings were confirmed in an experimental realization of a system with a single-particle mobility edge \cite{AL5}. There is a vast literature on mobility edges in Hermitian systems, but it has only been recently that mobility edges have been explored for various 1D tight-binding non-Hermitian models \cite{NH0,NH1,NH2,NH3,NH4,NH5,NH6,NH6a,NH7,NH8,NH9,NH10,NH11,NH13,NH14,NH15,NH16,NH16cemy,NH17,NH17jgek}. The first such model was considered in the pioneering paper by Hatano and Nelson \cite{hatano}. In non-Hermitian systems, in comparison to the Hermitian ones, the mobility edges not only separate localized states from the extended states but also indicate the coexistence of complex and real energies. The latter allows us to come out with a topological characterization of mobility edges \cite{NH1}. Apart from these models, extended and localized states can coexist in some other Hermitian lattices with inhomogeneous trap \cite{NH1a,NH1b} and with partially disordered potential \cite{NH1d}. In general, such systems require complicated engineering of the hopping parameters and onsite potentials \cite{NH1c}.

In this work we consider non-Hermitian extensions of the one dimensional Anderson and Aubry-Andr\'e-Harper models with asymmetric (nonreciprocal) hopping amplitudes at which non-Hermitian skin effect (NHSE) plays important roles on the localization \cite{cy0,cy1,cy2,ek1,cy3,cy4,cy5,ek4,cy6,cy7,cy8,ek2,ek3,ek5,ek6}. We introduce mixed boundary conditions (MBC) as a mixture of periodic (PBC ) and open (OBC) boundary conditions and show that extended and localized states can coexist even for the lattice without the disorder. We show that extended states form a closed loop in the complex energy plane while the localized states have real energies. We further explore the effect of onsite potentials and show that localized and extended states survive in the presence of the onsite potentials until topological phase transition occurs at strong disorder and all states are localized.

\section{Model}

The starting point of our analysis is provided by the one-dimensional nonreciprocal lattice with asymmetric nearest-neighbor couplings and onsite potentials. The field amplitudes $\ds{\psi_n}$ at various sites of the lattice can be obtained by solving
\begin{equation}\label{rof64oalk2} 
	J_R~\psi_{n-1}+J_L  ~\psi_{n+1} +V_n~ \psi_n  =E~\psi_n
\end{equation} 
where $\ds{n=1,2,,...,N}$ with  $\ds{N}$ being the total number of sites, $J_L$ and $J_R$ are positive-valued coupling constants in the left and right directions, respectively, $\ds{V_n}$ are real-valued onsite potentials. We assume $\ds{J_L>J_R}$, unless otherwise stated. Two different types of onsite potentials should be distinguished here. The first one is for the non-Hermitian Anderson model at which the onsite potentials are independent random potentials uniformly distributed in the interval $\ds{ W [ - \frac{1}{2} ,  \frac{1}{2}   ]  }$ with disorder strength $\ds{W}$. This model exhibits an Anderson transition at a non-zero value of the disorder strength in contrast to the Hermitian system, whose eigenstates are always localized in the presence of a random potential \cite{hatano,PRX}. The second one is for the non-Hermitian Aubry-Andr\'e model at which the onsite potential is the quasi-periodic potential to describe an intermediate case between ordered and disordered systems, i. e., $\ds{V_n=V_0 \cos{ (2\pi {\beta} n )} }$, where $\ds{V_0}$ is the amplitude of the onsite incommensurate potential and $\ds{ \beta }$ is an irrational number. This model exhibits a metal-insulator transition when the potential strength is above a critical point \cite{cy7}.

\begin{figure}[t]
\includegraphics[width=\columnwidth]{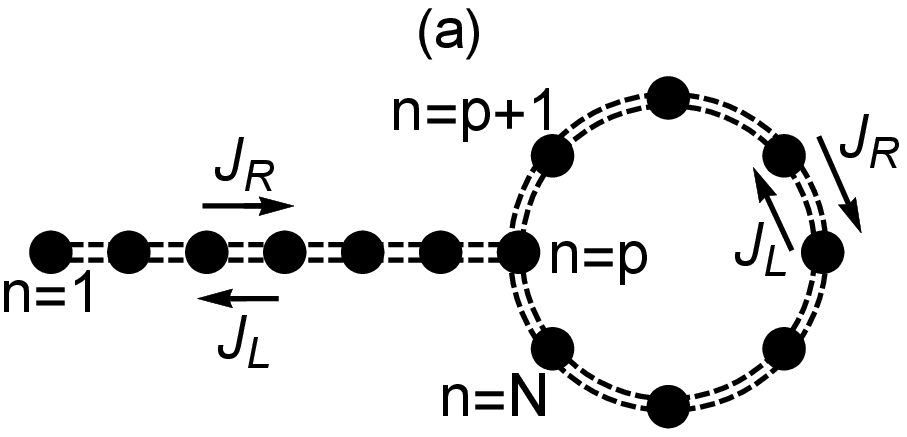}
\includegraphics[width=4.2cm]{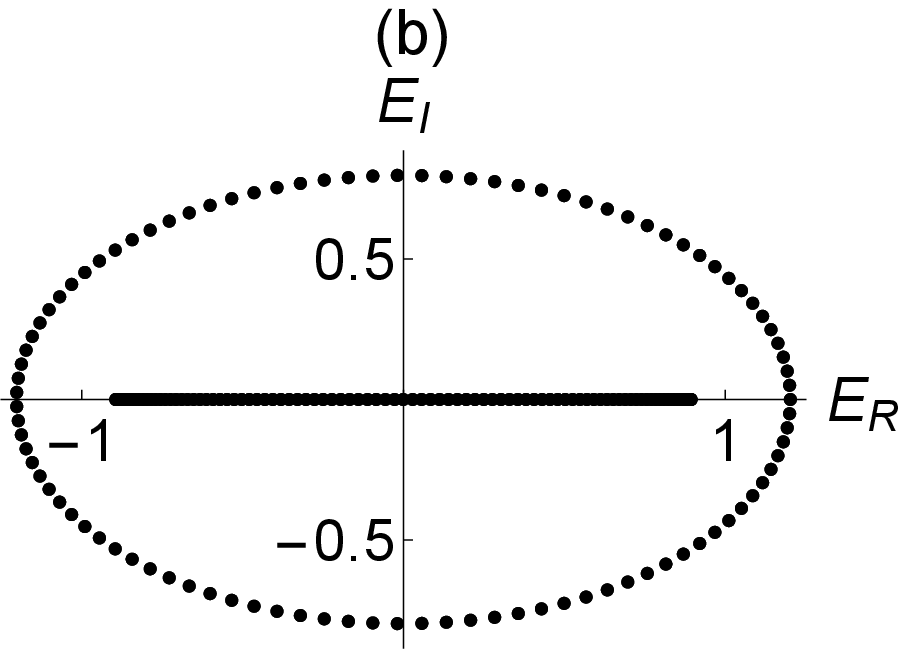}
\includegraphics[width=4.2cm]{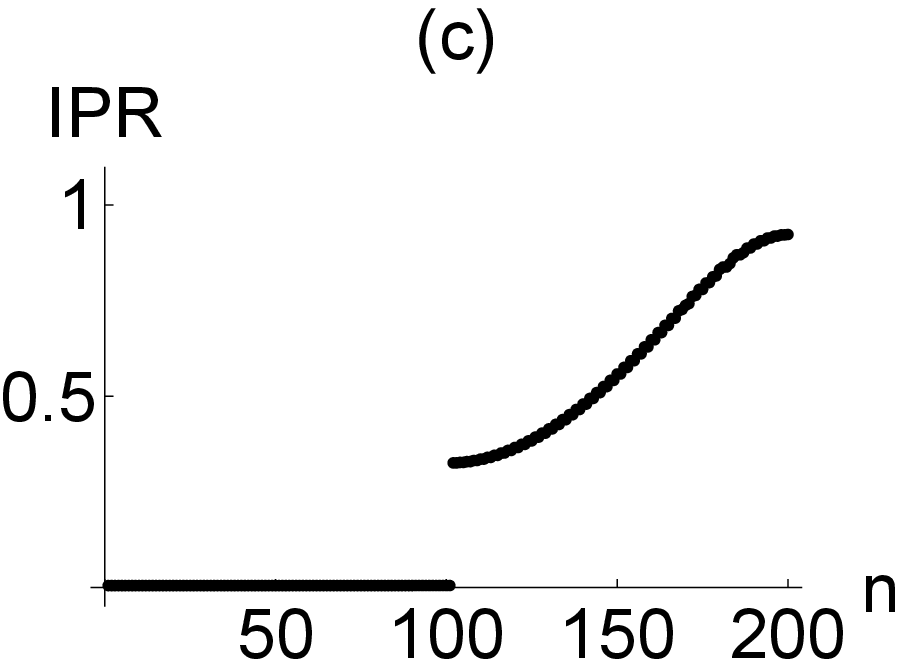}
\includegraphics[width=4.2cm]{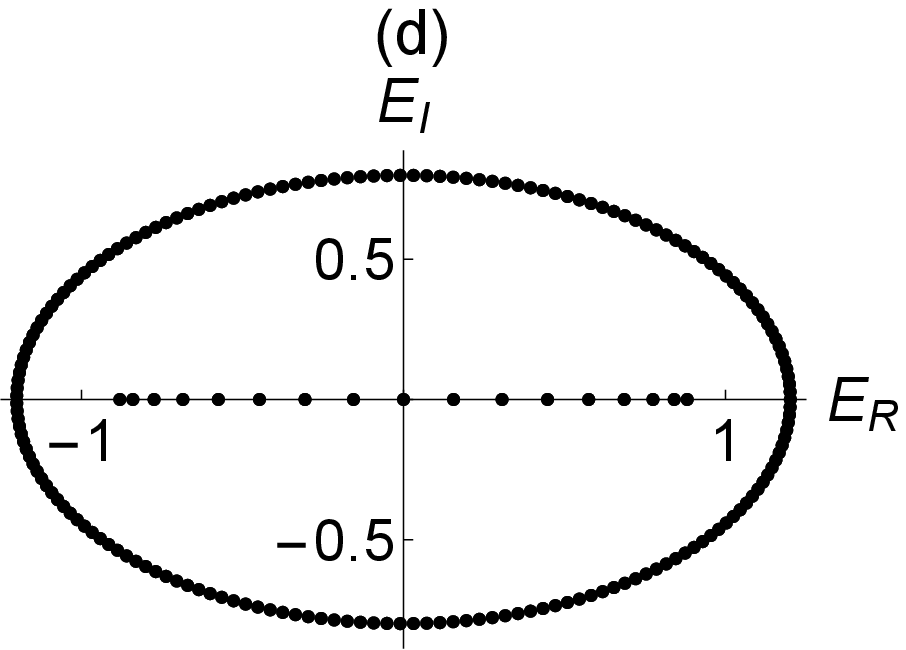}
\includegraphics[width=4.2cm]{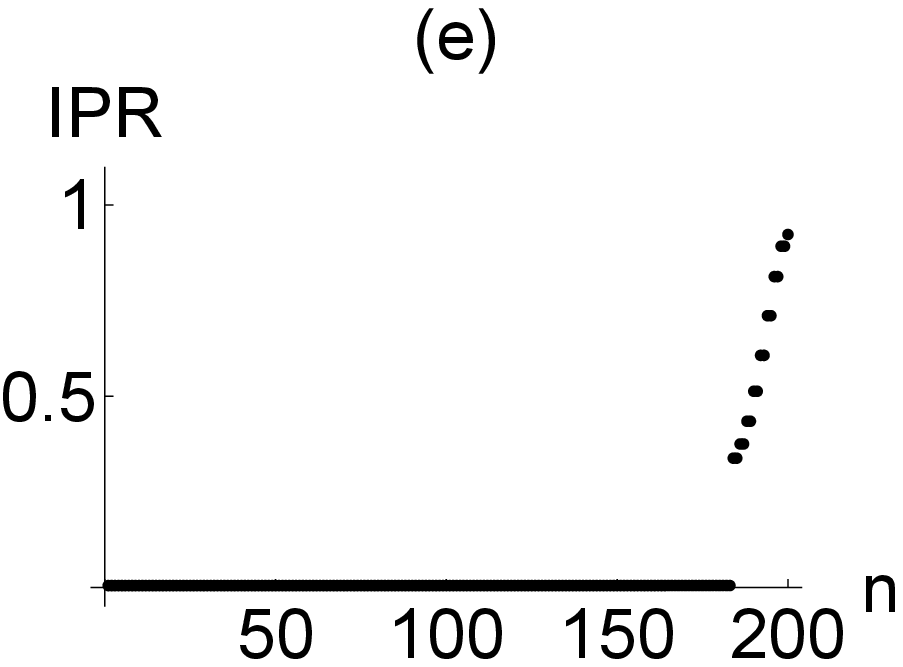}
\caption{ (a) A representation for a lattice with asymmetrical couplings under MBC with $\ds{ N=14 }$ and $\ds{p=7}$. The lattice has one edge and one circular ring. In the ring, the couplings are $\ds{J_R}$ in the clockwise direction and $\ds{J_L}$ in the counterclockwise direction. (b,d) The energy spectra in the complex plane, where extended states are placed on the loop and localized states are placed on the real axis inside the loop. (c,e) The sudden jump from almost zero to a large IPR values indicates the coexistence of extended and localized eigenstates. The numerical parameters are $\ds{J_L=1}$, $\ds{J_R=0.2}$, $\ds{N=200}$ and $\ds{p=100}$ (b,c) and $\ds{p=18}$ (d,e). The PBC (OBC) spectrum is denser for a (smaller) larger $p$. The total number of the states with almost zero IPR values are equal to $\ds{N-p}$. }
\label{fig1}
\end{figure}

The spectrum for the non-Hermitian lattice described by Eq. (\ref{rof64oalk2}) shows strong sensitivity to the boundary conditions in topologically nontrivial region \cite{PRX}. Consider for example, the case without onsite potentials, which is topologically nontrivial as long as $\ds{J_L{\neq}J_R}$. In this case, the spectrum describes a loop in the complex energy plane when the lattice has no edges (under PBC), whereas the spectrum is real when the lattice has two edges (under OBC). The change in the spectrum is also dramatic if the lattice has only one edge. In fact, there can be two such cases. The first one is the semi-infinite lattice ($\ds{N \rightarrow\infty}$) whose spectrum fills the interior of the PBC loop in the complex plane. However, this case is not physical since any experiment naturally contains a finite number of lattice sites. The second one is the finite lattice with only one edge, i. e., the lattice has an open edge on the left and the other edge is bent to form a circular ring on the right. Suppose that the right end of the lattice is coupled to the lattice at the lattice site $p$. For an illustration, such a lattice with $\ds{N=}14$ and $\ds{p=7}$ is depicted in Fig.\ref{fig1} (a). In this case, the system satisfies mixed boundary conditions (MBC). In this case, Eq. (\ref{rof64oalk2}) is modified at $n=p$ (due to the extra coupling at $n=p$)
%\begin{eqnarray}\label{mbccydwq} 
%	J_R~\psi_{n-1}+J_L  ~\psi_{n+1} +V_n~ \psi_n & =& E~\psi_n ~ (n{\neq} p  ) \nonumber\\
%		J_R~(\psi_{p-1}+ \psi_{N})+J_L  ~\psi_{p+1} +V_p~ \psi_p  &=&E~\psi_p 
%\end{eqnarray}
\begin{equation}
\begin{array}{cc}
J_R\psi_{n-1}+J_L \psi_{n+1} +V_n\psi_n  =E\psi_n &(n{\neq} p  ) \\
J_R(\psi_{p-1}+ \psi_{N})+J_L \psi_{p+1} +V_p\psi_p  =E\psi_p &(n= p  )
\end{array}
\end{equation}
where $\ds{p}$ is a site number in the bulk $\ds{ 2~{\leq}~p~{\leq}~N-1 }$. Note that in order to obtain the solution of the former equation, we suppose 
\begin{equation}\label{mbccy} 
\psi_{0}=0~,~~~~~\psi_{N+1}=\psi_p
\end{equation} 
In the Hermitian lattice, $J_L=J_R$, MBC is of no special importance since the extra coupling between the right edge and a bulk point of the lattice has only perturbative effects for a long lattice (the MBC, PBC and OBC energy spectra almost coincide). On the other hand, in the non-Hermitian lattice, MBC leads to the coexistence of extended (delocalized) and localized eigenstates even in the absence of any onsite potentials. We emphasize that the delocalized states are not extended only in the circular ring, but throughout the whole lattice. Note that such a coexistence was shown to appear in the presence of tailored quasi-periodical potentials and coupling constant \cite{NH1,NH2,NH3,NH4,NH5,NH6,NH6a,NH7,NH8,NH9,NH10,NH11,NH13,NH14}. However, we see it in our system as a result of the boundary condition sensitivity of the nonreciprocal non-Hermitian systems. 

Let us start with the case without onsite potentials, $\ds{V_n=0}$ in a long but finite lattice. The spectrum under MBC describes both a line segment on the real axis and a loop that is slightly deformed from the PBC loop in the complex plane. Nat. Comm. 11, 5491 2020The states distributed on the MBC loop are extended states, whereas the ones on the line segment are skin states that are exponentially localized at the left edge. The parameter $\ds{p}$ has the key role on the total number of extended states. In fact, there are $\ds{N-p+1}$ extended eigenstates and the rest are all skin states. As a special case, we have only one skin state that is also topologically robust against the coupling disorder at $\ds{p=2}$. Oppositely, at $\ds{p=N-1}$, there exists one pair of extended states $\ds{ \{  \psi_n ,e^{i{\pi}  n}  ~\psi_n \} }$ with real energies and all other states are localized skin states. To quantify localization and extension of an eigenstate with eigenvalue $\ds{E}$, we can use the inverse participation ratio (IPR) 
\begin{equation}\label{ipr} 
	IPR(E)=  \frac{     \sum_n |\psi_n(E)  |^4 }{ ( \sum_n|\psi_n  (E)  |^2)^2 }
\end{equation}
Specifically, IPR is of the order of $\ds{1/N}$ for an extended eigenstate while it is close to $\ds{1}$ for a localized eigenstate. To illustrate our discussion, we firstly plot the spectra in the complex plane for two different values of $p$ at $J_L=1$, $J_R=0.2$ and $N=200$ in Fig.\ref{fig1} (b,d). The points on the loop are very dense for small values of $\ds{p}$ and become sparse with increasing $\ds{p }$ at fixed $\ds{N}$. The line segment on the real axis is always in the MBC loop. We then plot the IPR values corresponding to the cases (b,d) in Fig.\ref{fig1} (c,e). One can notice the gap in these plots where the IPR values jump from almost zero values to nearly $0.4$ at $n=p-1$. This sharp increase of IPR implies the coexistence of localized and extended states in the absence of the disorder. To this end, let us write the analytical solution available for the unidirectional lattice with $\ds{J_R=0}$ under MBC. In this case, the extended state is given by $\ds{\psi_n= e^{ik n} }$ with eigenvalues $\ds{    E=J_L~e^{ik} }$, where $\ds{k={ 2{\pi }j}(N-p+1)^{-1} }$ and $\ds  {j=0,1,...,N-p}  $, respectively. There is just one skin state $\ds{\psi_n= \delta_{n,1} }$ at zero energy since the system has an exceptional point of order $\ds{ N-p   }$. 

\begin{figure}[t]
\includegraphics[width=4.25cm]{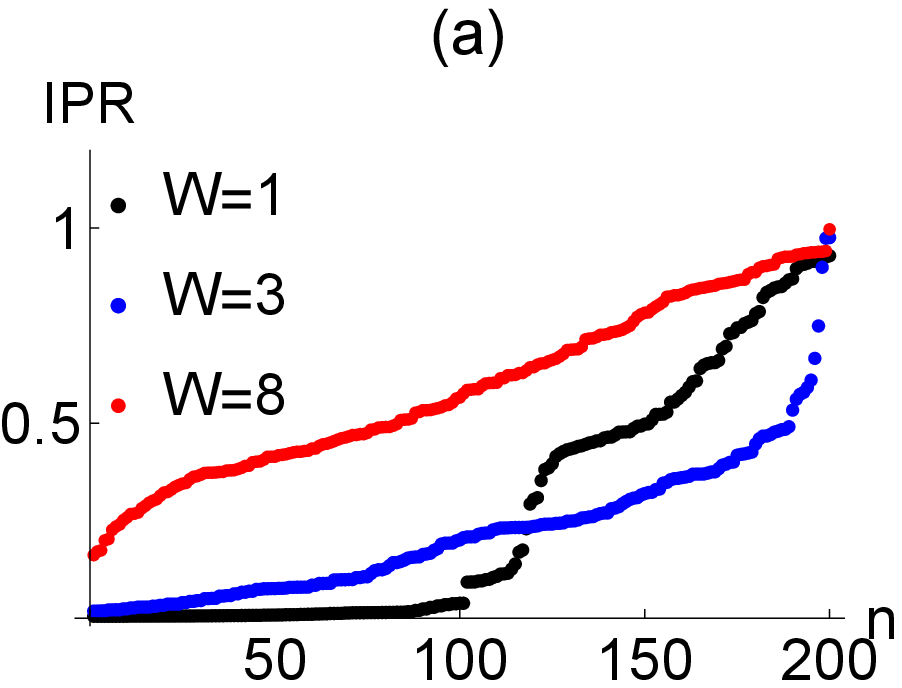}
\includegraphics[width=4.25cm]{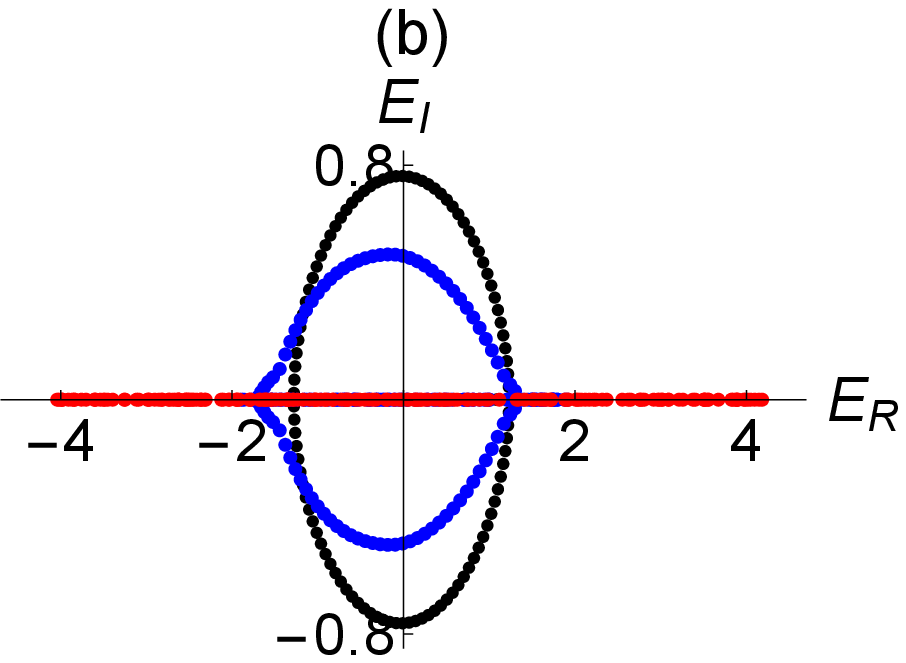}
\includegraphics[width=4.25cm]{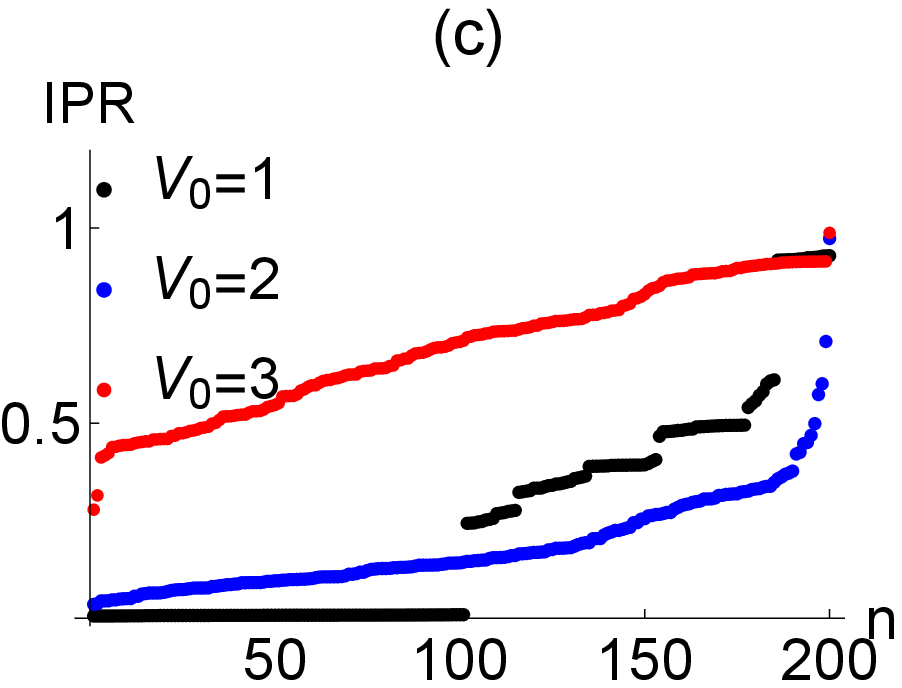}
\includegraphics[width=4.25cm]{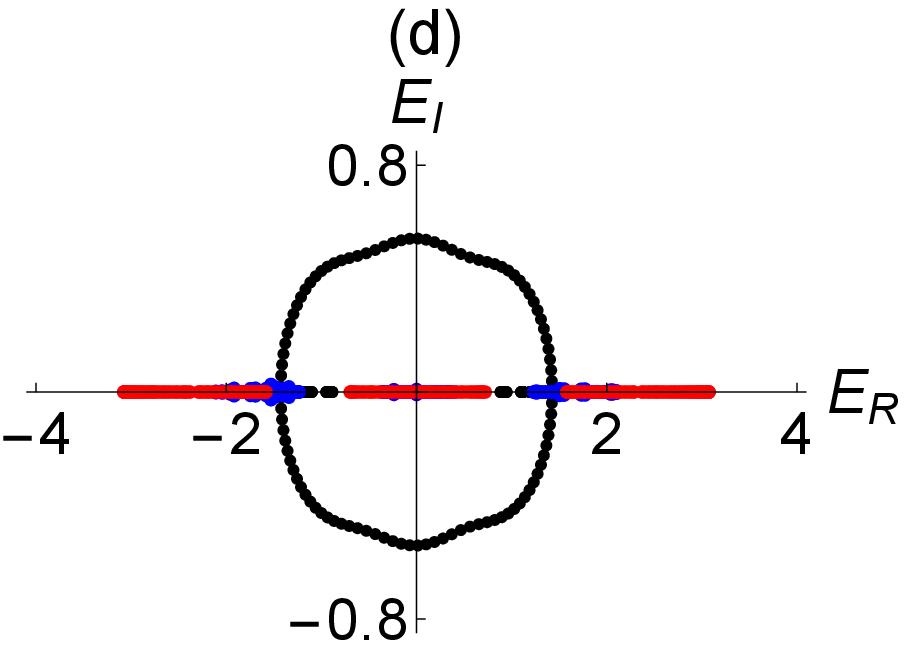}
\caption{ IPR values and their corresponding energy eigenvalues in the complex plane at various potential strengths for Anderson (a,b) and Aubry-Andr\'e models (c,d), respectively,  when $\ds{J_L=1 , J_R=0.2}$ and $\ds{N=2p=200}$. At strong onsite potentials (in red), all eigenvalues lie on the real axis, indicating that all eigenstates are localized. Contrarily, at weak onsite potentials (in black), almost half of the eigenstates are extended while the other half are localized. In the intermediate case (blue), there are still a few extended eigenstates. Note that $V_0=2$ is the phase transition point for the quasi-periodical potential. One can see a few extended states with small complex eigenvalues due to the finite number of the lattice sites (localization length is large and it practically becomes extended. If the lattice is much longer, then one would see its localization character).}
\label{fig2}
\end{figure}

Introducing disorder through random onsite potentials deforms the energy loop in the complex plane at fixed $\ds{p}$ (contraction in the imaginary axis and elongation in the real axis as the disorder strength increases). Furthermore, it reduces the total number of extended states described by the points on the energy loop and hence increases the total number of localized states described by the points located on the real axis. At weak disorder strength, localized states are mostly skin states localized at the left edge. Beyond the Anderson transition point at which all eigenstates are localized, localization occurs all over the lattice. We plot the IPR values and complex energy spectra in Fig.\ref{fig2} (a,b) for the system described in Fig.\ref{fig1} but with various disorder strengths. As can be seen, increasing the disorder strength reduces the total number of extended states until the disorder strength is equal to a critical strength ($W_c\approx5$ ) at which Anderson transition occurs. Therefore, there are still some extended eigenstates at $W=1$ (in black) and $W=3$ (in blue), but all eigenstates are localized at $W=8$ (in red). The corresponding spectrum becomes real valued and the OBC and MBC spectra are almost the same when all eigenstates are localized (Fig.\ref{fig2} (b)). The Anderson transition point also corresponds to a topological phase transition point as we will see below. As a result, we say that extended and localized states coexist only in the topologically nontrivial region. The critical disorder strength at which Anderson transition occurs depends on $\ds{p}$ at fixed $N$. Roughly speaking, $\ds{W_c}$ at fixed $N$ increases slightly with $\ds{p}$ unless $p$ is close to $N$ at which $ \ds{W_c}$ decreases sharply since the system has already a few extended eigenstates even in the absence of the disorder. 

\begin{figure}[t]
\includegraphics[width=\columnwidth]{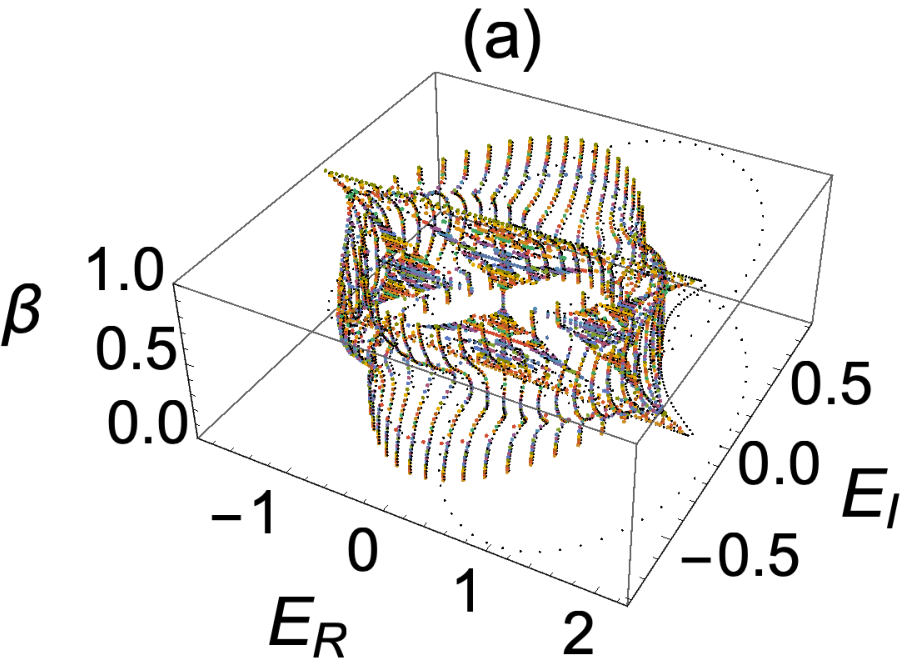}
\includegraphics[width=\columnwidth]{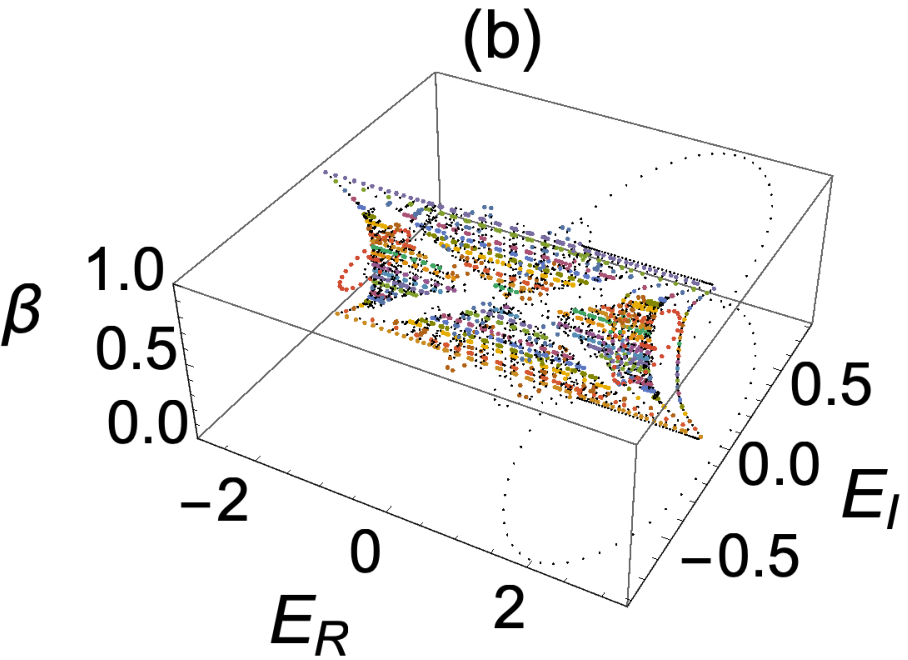}
\caption{ The butterfly spectra under MBC in 3 dimensions at $V_0=1$ (a) and $V_0=2$ (b), where $E_R$ and $E_I$ are real and complex parts of the energy eigenvalues. In 2 dimensional complex energy plane, the spectrum determines a loop and a line inside the loop as in Fig. 1 (b). In 3 dimensions where $\beta$ is the vertical axis, the butterfly shape appears. Note that such a three dimensional butterfly structure is not possible under both PBC and OBC. The parameters are given by $\ds{J_L=1}$ and $\ds{J_R=0.2}$ and $\ds{N=2p=100}$. }
\label{fig3}
\end{figure}

We perform another computations for the quasi-periodical potential $\ds{V_n=V_0 \cos{ (2\pi {\beta} n )} }$ and plot the IPR values and energy spectra in Fig.\ref{fig2} (c,d) and for three different values of $\ds{V_0}$ at $\ds{ \beta=\frac{\sqrt{5} -1}{2}}$ and $\ds{p=\frac{N}{2}}$. At $V_0=1$, we see a sharp increase in the IPR values from $0$ to nearly $0.3$, indicating that almost half of the states are extended while the rest are localized (in black). It is well known that the critical point at which localization-delocalization transition occurs is at $2$ in the Hermitian Aubry-Andre model. This value is almost equal to the critical point for the MBC (a slight perturbation comes from the left edge and coupling between the right edge and the lattice point $p$). The critical point also coincides with the topological phase transition point as we will see below. One can see a few complex eigenvalues (in blue) at $V_0=2$ with complex eigenvalues in Fig.\ref{fig2} (d) (in blue). Beyond the critical point the spectrum is real and all eigenstates are localized ( $V_0=3$ in red). As a result, we say that extended and localized states coexist in the quasi-periodical lattice under the MBC as long as $V_0$ is below than the critical number at which a topological phase transition occurs. To this end, we plot the butterfly spectra for the MBC in 3 dimensions, where $\ds{\beta}$ is the vertical axis and the other two axes are the real and imaginary parts of the spectrum (Fig.\ref{fig3} ). Note that such a three dimensional butterfly structure is not possible under OBC, which has real spectrum.

Let us discuss topological features in our system. The spectral winding number $\omega$ at the zero base energy for the Hatano-Nelson model in the absence of onsite potentials is equal to $\ds{-1  }$ when $\ds{  J_L>J_R }$ \cite{PRX}. The system remains to be in the topological phase in the presence of onsite disorder until the disorder strength is strong enough to make all eigenstates to have real eigenvalues at which the Anderson transition occurs. To compute the topological number in the presence of the onsite potentials under MBC, we follow a similar method introduced in Ref. \cite{PRX}. Suppose that the coupling constant at the lattice closing point (between $N^{\text{th}}$ and $p^{\text{th}}$ sites) are multiplied by $e^{{\mp} i \Phi   }$, where $\ds{\Phi}$ is a fictitious magnetic flux. Then the winding number at zero base energy for a disordered lattice is given by
\begin{figure}[t]
\includegraphics[width=4.25cm]{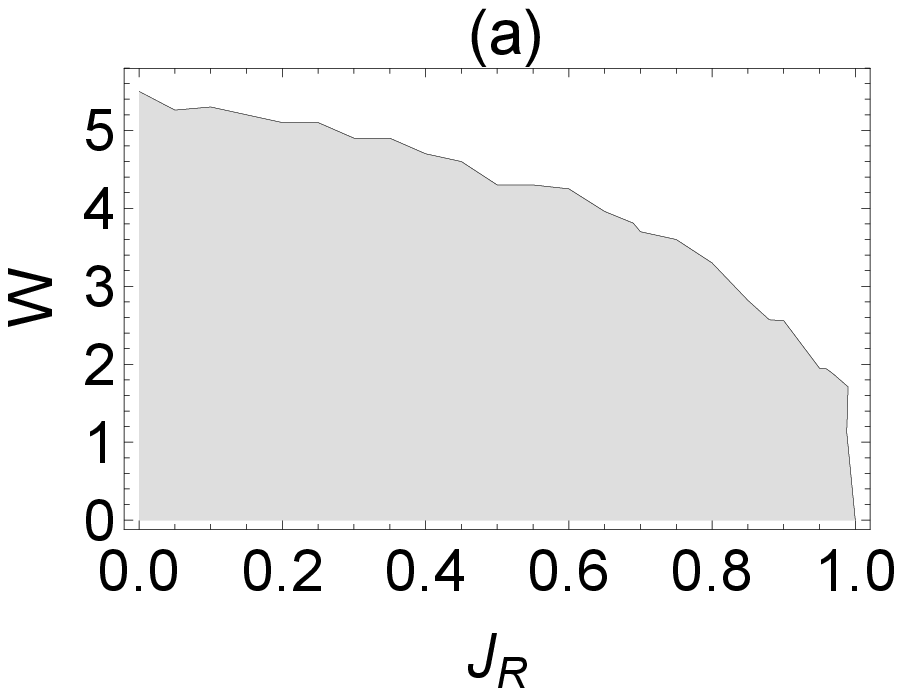}
\includegraphics[width=4.25cm]{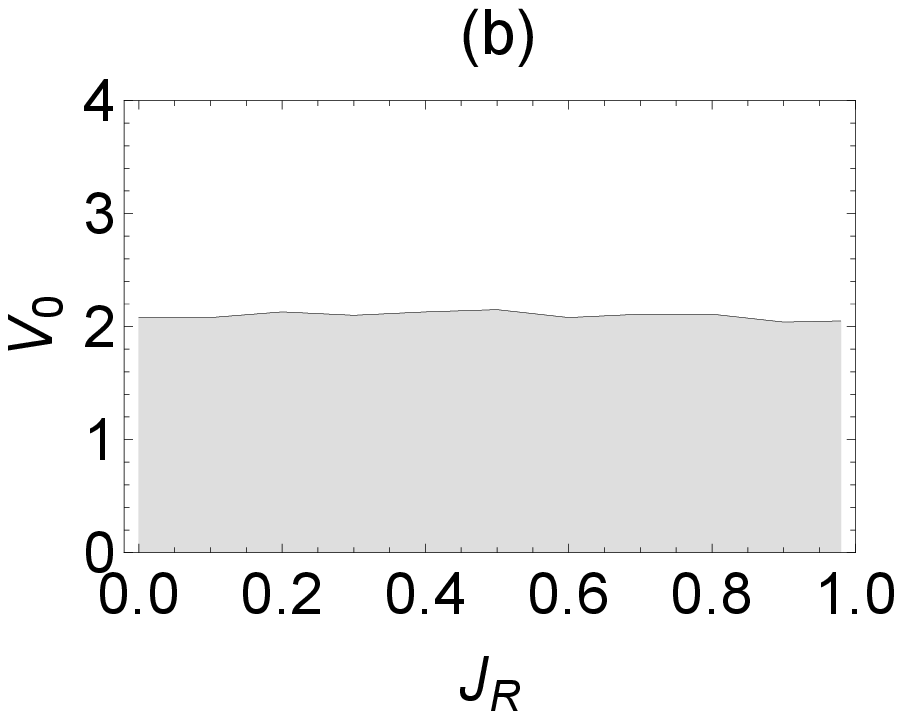}
\caption{ The critical strengths for the Anderson (a) and Aubry-Andre (b) models under MBC as a function of $J_R$ at $J_L=1$ and $N=2p=200$. The shaded are has winding number $\omega=-1$ and complex spectra. The top unshaded are has zero winding number and real spectra. In topologically nontrivial region (shaded area), localized and extended states coexist. On the other hand, in topologically trivial region, only localized states exist. }
\label{fig4}
\end{figure}
\begin{equation}\label{73027fhk2} 
	\omega=\int_0^{2\pi} \frac{ d\Phi }{2 {\pi} i } \partial_{\Phi} ln \det(H(\Phi))
\end{equation} 
where $\ds{H}$ is the corresponding Hamiltonian for the model (\ref{rof64oalk2}) under MBC. The spectral winding number counts the number of times the complex spectral trajectory encircles $E_B=0$ base energy when $\ds{\Phi}$ varies from zero to $\ds{2\pi}$. Apparently, the winding number becomes zero when the spectrum is real and all eigenstates are localized. Note that the above formula works well when the number $p$ is not close to $N$ since the spectral loop in the complex plane is less denser when $p$ increases. The MBC lattice is required to be a finite lattice, so we approximate the derivative with finite difference in the numerical differentiation. We present our numerical results and plot the winding number as a function of $J_R$ in Fig.\ref{fig4}, where the shaded and unshaded area has $w=-1$ and $w=0$, respectively. In (a), the critical strength is around $W\approx5$ at $J_R=0$ and reduced to zero at $J_R=1$ (the spectrum becomes real in the Hermitian limit). On the other hand, it is almost constant for the quasi-periodical lattice (b). Small fluctuations around $V_0=2$ is the result of the perturbative effect due to the imposition of the MBC on the finite lattice.\\

We finally make a brief discussion for $\ds{ J_R>J_L }$. Without loss of generality, we suppose that $J_R=1$. Consider first that $V_n=0$. Due to NHSE, bulk states are localized at the right edge under OBC. If we consider the MBC, the right edge is coupled to a bulk point. In this case, there are $\ds{N-p+1}$ extended states and the rest are exponentially localized states centered at the bulk point $p$ where the right edge is closed. As opposed to the cases considered above, the extended states are extended only in the circular lattice at any value of $J_R$ and localized states have complex eigenvalues. Therefore, there are multiple energy loops in the complex plane, one for the extended states and another one(s) for the localized states. The localization length of the localized state increases and diffuses more into the straight lattice ($\ds{ n{\leq}p}$) as $J_R$ is increased. In the presence of the disorder, the number of extended states decreases and localized states appear centered at various points of the lattice. If the disorder is sufficiently strong, then Anderson transition takes place and all eigenstates have real eigenvalues and get localized.

\section{Conclusion}

It is generally believed that mobility edges separating extended and localized states in one-dimensional tight-binding models appear if correlated disorder and coupling constants are specially tailored. Here we introduce the mixed boundary conditions to study a finite lattice with one open edge as an alternative to the semi-infinite boundary conditions, which also requires one open edge. The finite lattice we consider is the one whose one edge is bent to form a circular ring and coupled to the lattice at the lattice point $p$. We have shown that extended and localized states can coexist even without onsite potentials in such a lattice as a result of the boundary condition sensitivity of the nonreciprocal non-Hermitian systems as long as the system is topologically nontrivial. We have also shown that the total number of extended states is exactly equal to $N-p+1$, where $N$ is the total number of the lattice sites. In the presence of the disordered onsite potentials, the total number of the extended states reduces with increasing disorder strength and the extended states disappear when the disorder strength is at the critical point at which topological phase transition occurs since the corresponding spectrum become real valued. Experimental observation of mobility edges in non-Hermitian systems often requires complicated designs of couplings or onsite potentials. The mixed boundary conditions can be utilized in non-Hermitian systems to obtain mobility edges more easily. \\
{\it{Acknowledgments--}}	C. Y. wishes to acknowledge the support from the Scientific and Technological Research Council of Turkey through the 2219 program with grant number 1059B191900044. H. R. acknowledge the support by the Army Research Office Grant No. W911NF-20-1-0276 and NSF Grant No. PHY-2012172. The views and conclusions contained in this document are those of the authors and should not be interpreted as representing the official policies, either expressed or implied, of the Army Research Office or the U.S. Government. The U.S. Government is authorized to reproduce and distribute reprints for Government purposes notwithstanding any copyright notation herein.

\end{document}